\documentclass{ws-ijcm}
\usepackage[authoryear,semicolon]{natbib}
\let\textquotedbl="

\begin{document}

%%%%%%%%%%%%%%%%%%%%% Publisher's Area please ignore %%%%%%%%%%%%%%%
%
\catchline{}{}{}{}{}
%
%%%%%%%%%%%%%%%%%%%%%%%%%%%%%%%%%%%%%%%%%%%%%%%%%%%%%%%%%%%%%%%%%%%%

\markboth{A. Polanski, M. Marczyk, M. Pietrowska, P. Widlak, J. Polanska}
{Initializing EM by dynamic programming partitions}

\title{INITIALIZING EM ALGORITHM FOR UNIVARIATE GAUSSIAN, MULTI-COMPONENT, HETEROSCEDASTIC MIXTURE MODELS BY DYNAMIC PROGRAMMING PARTITIONS}

\author{ANDRZEJ POLANSKI}

\address{Institute of Informatics, Silesian University of Technology \\ 44-100 Gliwice, Poland \\
\email{andrzej.polanski@polsl.pl}}

\author{MICHAL MARCZYK}
\address{Data Mining Group, Institute of Automatic Control, Silesian University of Technology \\ 44-100 Gliwice, Poland \\
\email{michal.marczyk@polsl.pl}}

\author{MONIKA PIETROWSKA}
\address{Maria Sklodowska-Curie Memorial Cancer Center and Institute of Oncology, Branch in Gliwice \\ 44-101 Gliwice, Poland \\
\email{monika.pietrowska@io.gliwice.pl}}

\author{PIOTR WIDLAK}
\address{Maria Sklodowska-Curie Memorial Cancer Center and Institute of Oncology, Branch in Gliwice \\ 44-101 Gliwice, Poland \\
\email{widlak@io.gliwice.pl}}

\author{JOANNA POLANSKA}
\address{Data Mining Group, Institute of Automatic Control, Silesian University of Technology \\ 44-100 Gliwice, Poland \\
\email{joanna.polanska@polsl.pl}}

\maketitle

\begin{history}
\received{(Day Month Year)}
\revised{(Day Month Year)}
\end{history}

\begin{abstract}
Setting initial values of parameters of mixture distributions estimated by using the EM recursive algorithm is very important to the overall quality of estimation. None of the existing methods is suitable for mixtures with large number of components. We present a relevant methodology of estimating initial values of parameters of univariate, heteroscedastic Gaussian mixtures, on the basis of the dynamic programming algorithm for partitioning the range of observations into bins. We evaluate variants of dynamic programming method corresponding to different scoring functions for partitioning. For simulated and real datasets we demonstrate superior efficiency of the proposed method compared to existing techniques.
\end{abstract}


\section{Introduction}

\label{intro} Due to the importance of Gaussian distribution
a significant part of the research on
improving performance of the expectation maximization
(EM) recursive algorithm [\cite{McLachlan2000}] for estimation of 
parameters of mixture distributions is focused 
on estimating parameters of mixtures of Gaussian components.
A problem of crucial importance is the~choice of~initial values for
mixture parameters. A~good choice of a starting point for the EM
iterations can result in reducing the probability of erroneous estimation
of parameters and/or in (faster) convergence of EM iterations. Approaches
to initializing EM iterations were extensively discussed and studied [\cite{McLachlan2000, Karlis2003, Biernacki2003, Biernacki2004, Maitra2009, OHagan2012,
Melnykov2012}]. A~simple approach is random initialization
involving generation of initial values of parameters and component
weights (mixing proportions) by using some assumed probability distributions
[\cite{McLachlan2000}]. Another simple idea is using data quantilles
to estimate initial means and variances of components to start EM
iterations. A group of approaches involve using some kind of clustering
procedure (hierarchical clustering or k-means clustering) applied
for the data set to compute initial parameters for EM iterations [\cite{Biernacki2003,
Maitra2009, OHagan2012}]. Some other
ideas for computing initial values involve using sample moments of
data [\cite{Karlis2003}], method of using singular value decompositions
for multivariate mixtures [\cite{Melnykov2012}], using properties
of EM trajectories and data moments to limit the search space for
initial parameters of multivariable Gaussian mixtures [\cite{Biernacki2004}].
Available software packages for mixture modeling [\cite{McLachlan1999,
Biernacki2006, Fraley1999, Richardson1997}] offer different possibilities for setting
initial conditions for EM iterations.

A challenge for the existing methodologies 
is decomposition of univariate, heteroscedastic Gaussian
mixtures with large number of components. It can be verified in practical
computations that with the increase of the
number of components in the mixture, the application of the EM algorithm
with the published methods for setting initial conditions would lead
to the mixture models of the progresively worse quality. Yet, problems
of mixture decompositions of univariate models 
where the numbers of components are large
are encountered in many applications. Some examples are given hereafter.
Frequency (amplitude or power) spectra of different time domain signals
may contain numerous compenents. E.g., in the vibration diagnostics
or speech recognition applications frequency spectra of measurement
signals can be analyzed as mixtures of tens of Gaussian
functions [\cite{Machado2013}]. In nuclear magnetic resonance (NMR)
spectroscopy free induction decay (FID) signal contains components
corresponding to molecules present in the sample, associated
with different frequencies. Frequency spectrum of such signal contains
tens of components. Moreover, spectral signatures of some metabolites
can exhibit complex shapes, which may require including even more
components for their modeling [\cite{Chylla2012}]. In some applications
concerning modeling, interpolation and forecasting of time signals
mixtures of Gaussian functions are applied, which
include tens of components [\cite{Eirola2013}]. Time of flight
mass spectrometry, a high-throughput, experimental technique in molecular
biology provides measurements of peptide, protein or lipid compositions
of biological samples. Mass spectral signals can be modeled by mixture
decompositions, where numbers of components can reach even several
hundreds [\cite{Pietrowska2011, Polanski2015}].

The aim of this paper is to develop and evaluate the method for estimating
initial values of parameters for EM iterations, for univariate, multi-component,
heteroscedastic Gaussian mixtures, based on the dynamic programming
partitioning algorithm. Partitioning the data points into bins, by
dynamic programming, determines initial values of weights, means and
variances of Gaussian components for the EM algorithm. The idea of
partitioning an interval into bins, by using dynamic programming,
with the aim of obtaining solution to some fitting or estimation problem
was formulated by Bellman [\cite{Bellman1961}] and then studied/used by other authors
in many contexts [e.g., \cite{Hebrail2010}]. The advantage of using
dynamic programming over partitioning (clustering) heuristics by hierarchical
clustering or greedy algorithms is that the dynamic programming method
allows for obtaining global optimal solution for a given quality function.
According to authors' best knowledge, dynamic programming partitions
were so far not used for initalization of EM iterations.

We present the dynamic programming algorithm for computing initial
values for mixture parameters for EM iterations. We also study the
problem of the choice of the scoring function. We compare several
possible scoring functions and, on the basis of computational experiments,
we propose the one best suited to the problem of computing initial
values of mixture parameters. For a number of artificially generated
univariate data sets, with multiple Gaussian, heteroscedastic components
we compare some other methodologies of initialization of EM iterations
with the dynamic programming method and we show its advantage. In
the areas of applications of mixture decompositions listed afore,
measured signals are often numbers of counts or intensities. In the
context of the mixture decompositions of such datasets we describe
the version of EM iterations appropriate for binned data and we develop
an approach for the dynamic programming partition for computing initial
conditions. We apply the dynamic programming method for initializing
EM iterations to the real dataset coming from protein mass spectrometry
experiment and we again demonstrate the efficiency of dynamic programming
method for initialization of the EM iterations.

\section{Gaussian mixture modeling}
Gaussian mixture modeling of the dataset of univariate scalar observations,
$\mathbf{x}=[x_{1},x_{2},...,x_{N}]$ involves estimating the vector
of mixture parameters 
\[
\mathbf{p}=[\alpha_{1},...,\alpha_{K},\mu_{1},...,\mu_{K},\sigma_{1},...,\sigma_{K}]
\]
(component weights, means and standard deviations) of the mixture
probability density function (pdf) 
\begin{equation}
f^{mix}(x,\mathbf{p})=\sum_{k=1}^{K}\alpha_{k}f(x,\mu_{k},\sigma_{k}),\label{G-mixture-model}
\end{equation}
where $f(x,\mu_{k},\sigma_{k})$ is the Gaussian pdf and $\sum_{k=1}^{K}\alpha_{k}=1$,
such that the log-likelihood function

\begin{equation}
L\left(\mathbf{x},\mathbf{p}\right)=\sum_{n=1}^{N}\log f^{mix}\left(x_{n},\mathbf{p}\right)\label{log-lik}
\end{equation}
is maximized [\cite{McLachlan2000}].

Analyzes of datasets where measurements are numbers of counts or
intensities involve formulations of mixture model decompositions for
binned data [\cite{McLachlan2000}], where $\mathbf{x}$ is a vector
of equally spaced centers of bins and observations are given by a
vector $\mathbf{y}=[y_{1},y_{2},...,y_{N}]$ of counts $y_{n},\:n=1...N$
generated by multinomial distribution with probabilities $p_{n}$
defined by areas of bins, 
\begin{equation}
p_{n}=\sum_{k=1}^{K}\alpha_{k}\left[\varPhi(x_{n}+\frac{\delta}{2},\mu_{k},\sigma_{k})-\varPhi(x_{n}-\frac{\delta}{2},\mu_{k},\sigma_{k})\right].\label{prob-n}
\end{equation}
In the above $\varPhi(x,\mu_{k},\sigma_{k})$ is the cumulative probability
distribution function of the Gaussian distribution and $\delta$ is
a bin width. We assume that bins are \textquotedbl{}dense\textquotedbl{},
which allows for approximating the area of the bin by the product
of its width and the probability density function at the bin center,
and consequently for using the log-likelihood function defined by the following formula 
\begin{equation}
L\left(\mathbf{x},\mathbf{p}\right)=\sum_{n=1}^{N}y_{n}\log f^{mix}\left(x_{n},\mathbf{p}\right).\label{log-lik-aprx}
\end{equation}

\subsection{EM iterations}
Maximization of the log-likelihood functions (\ref{log-lik}) and
(\ref{log-lik-aprx}) is done by using EM iterations - the successive
updates of parameter vector $\mathbf{p}^{old}\leftarrow\mathbf{p}^{new}\mathrm{,}$
where 
\[
\mathbf{p}^{old}=[\alpha_{1}^{old},...,\alpha_{K}^{old},\mu_{1}^{old},...,\mu_{K}^{old},\sigma_{1}^{old},...,\sigma_{K}^{old}],
\]
and 
\[
\mathbf{p}^{new}=[\alpha_{1}^{new},...,\alpha_{K}^{new},\mu_{1}^{new},...,\mu_{K}^{new},\sigma_{1}^{new},...,\sigma_{K}^{new}].
\]
Standard formulae for EM iterations are e.g., [\cite{McLachlan2000,
Bilmes1998}]. The version of EM iterations appropriate
for binned data with dense bins is similar to standard EM iterations.
It involves defining conditional distributions of hidden variables,
$\chi_{n}$ (corresponding to unknown assignments of observations
to components) 
\begin{equation}
P[\chi_{n}=k\left\vert x_{n}\right.]=\frac{\alpha_{k}^{old}f(x_{n},\mu_{k}^{old},\sigma_{k}^{old})}{\sum\limits _{\varkappa=1}^{K}\alpha_{\varkappa}^{old}f(x_{n},\mu_{\varkappa}^{old},\sigma_{\varkappa}^{old})}
\label{cond-dist}
\end{equation}
and updates of parameters estimates given by 
\begin{equation}
\alpha_{k}^{new}=\frac{\sum\limits _{n=1}^{N}y_{n}P[\chi_{n}=k\left\vert x_{n}\right.]}{\sum\limits _{n=1}^{N}y_{n}},\label{alpha-update-y}
\end{equation}
\begin{equation}
\mu_{k}^{new}=\frac{\sum\limits _{n=1}^{N}y_{n}x_{n}P[\chi_{n}=k\left\vert x_{n}\right.]}{\sum\limits _{n=1}^{N}y_{n}P[\chi_{n}=k\left\vert x_{n}\right.]},\label{mu-update-y}
\end{equation}

\begin{equation}
(\sigma_{k}^{new})^{2}=\frac{\sum\limits _{n=1}^{N}y_{n}(x_{n}-\mu_{k}^{new})^{2}P[\chi_{n}=k\left\vert x_{n}\right.]}{\sum\limits _{n=1}^{N}y_{n}P[\chi_{n}=k\left\vert x_{n}\right.]},\label{sig-update-y}
\end{equation}
$k=1,2,...K$.

\subsection{Preventing divergence of EM iterations}
Some assumptions should be made concerning execution of the EM iterations.
In the case of unequal variances of components of the Gaussian mixture,
considered here, the log-likelihood (\ref{log-lik}) or (\ref{log-lik-aprx})
is unbounded [\cite{Kiefer1956}]. Unboundedness results in a
possibility of encountering divergence of EM iterations in practical
computations and in the need for using approaches for preventing divergence
[\cite{Yao2010, Ingrassia2004}]. Here we prevent divergence of EM iterations
by a simple constraint conditions. Namely we do not allow
standard deviations of Gaussian components and mixing proportions
to fall below given threshold values $\sigma_{\min}$ and $\alpha_{\min}$
i.e., we augment equations for iterations for standard deviations
and component weights by additional constraints 
\begin{equation}
\sigma_{k}^{new}\leftarrow\max(\sigma_{k}^{new},\sigma_{\min})\label{sig-limit}
\end{equation}
and 
\begin{equation}
\alpha_{k}^{new}\leftarrow\max(\alpha_{k}^{new},\alpha_{\min}).\label{alpha-limit}
\end{equation}
The above constraints are sufficient to prevent divergence of EM iterations.

The constraints values assumed in EM iterations are $\sigma_{\min}=10^{-2},\alpha_{\min}=10^{-4}$
for the simulated datasets and $\sigma_{\min}=1,\alpha_{\min}=10^{-5}$
for the proteomic dataset.

\section{Problem formulation}
\label{Probform} The problem studied in this paper concerns determining
initial values for mixture parameters for EM iterations, $\mu_{k}^{\mathtt{ini}},\:\sigma_{k}^{\mathtt{ini}},\:\alpha_{k}^{\mathtt{ini}},\:k=1,2,...,K$,
for the best quality of the mixture parameters estimates. All methods
for setting initial conditions for EM iterations, studied and compared
here, rely on partitions of the data range performed according to
some criterion. We assume that observations $x_{1},x_{2},...,x_{N}$
are sorted in the ascending order 
\begin{equation}
x_{1}<x_{2}<...<x_{N}.\label{observations-sorted}
\end{equation}
Partitions are defined by blocks $B_{1},B_{2},...,B_{K}$, $k$-th
block contains samples $i,i+1,...,j$, $B_{k}=\{i,i+1,...,j\}$.

Partitions defined by blocks are used for computing initial values
for parameters. Initial means are computed as 
\begin{equation}
\mu_{k}^{\mathtt{ini}}=\frac{1}{j-i+1}\text{\ensuremath{\sum_{n=i}^{n=j}}}x_{n},\label{pmean}
\end{equation}
initial values for standard deviations are computed as 
\begin{equation}
\sigma_{k}^{\mathtt{ini}}=\sqrt{\text{\ensuremath{\frac{1}{j-i+1}\text{\ensuremath{\sum_{n=i}^{n=j}}}}}(x_{n}-\mu_{k}^{\mathtt{ini}})^{2}},\label{pstd}
\end{equation}
and initial values for mixing proportions are computed by 
\begin{equation}
\alpha_{k}^{\mathtt{ini}}=\frac{\#B_{k}}{N}=\frac{j-i+1}{N}.\label{ppart}
\end{equation}
In the above expression $\#B_{k}$ denotes the number of measurements in the block $B_{k}$.

For the case of the binned data the appropriate expressions for initial
values of parameters, implied by partitions given by blocks $B_{k}=\{i,i+1,...,j\}$,
are as follows. Initial values for means are computed as 
\begin{equation}
\mu_{k}^{\mathtt{ini}}=\sum_{n=i}^{n=j}x_{n}w_{n},\label{pmeanw}
\end{equation}
initial values for standard deviations are computed as 
\begin{equation}
\sigma_{k}^{\mathtt{ini}}=\sqrt{\sum_{n=i}^{n=j}w_{n}(x_{n}-\mu_{k}^{\mathtt{ini}})^{2}},\label{pstdw}
\end{equation}
and initial values for component weights are computed as 
\begin{equation}
\alpha_{k}^{\mathtt{ini}}=\frac{\sum_{\nu=i}^{\nu=j}y_{\nu}}{\sum_{n=1}^{N}y_{n}}.\label{ppartw}
\end{equation}
In expressions (\ref{pmeanw})-(\ref{ppartw}) by $w_{n}$ we denote
\begin{equation}
w_{n}=\frac{y_{n}}{\sum_{\nu=i}^{\nu=j}y_{\nu}}.\label{defw}
\end{equation}

\section{Initializing EM iterations by using dynamic programming partitioning
algorithm}
\label{EMdynpro} In this section we describe the algorithm for computing
partitions by using the dynamic programming method. Partitioning of
the observations involves defining $K$ blocks 
\begin{equation}
B_{1},B_{2},...,B_{K}\label{blocks}
\end{equation}
where (as already stated) each of the blocks is a range of successive
indexes of observations 
\begin{equation}
B_{k}=\{i,i+1,...,j\}.\label{b-range}
\end{equation}
For each of the blocks (more precisely, for data in the block) we
compute a scoring function, denoted either by $Q(B_{k})$
or by $Q(x_{i},x_{i+1},...,x_{j})$, 
\begin{equation}
Q(B_{k})=Q(x_{i},x_{i+1},...,x_{j}),\label{quality-index}
\end{equation}
The problem of optimal partitioning involves defining blocks (\ref{blocks}),
such that the cumulative scoring index $Q(B_{1},B_{2},...,B_{K})$
is minimized, 
\begin{equation}
Q(B_{1},B_{2},...,B_{K})=\sum\limits _{k=1}^{K}Q(B_{k})\rightarrow\min.\label{cumulated-quality}
\end{equation}

The solution to the optimal partitioning problem (\ref{cumulated-quality})
by dynamic programming [\cite{Bellman1961}] is obtained by iterative application
of the following Bellman equation 
\begin{equation}
Q_{1..j}^{opt}(k+1)=\min_{i=1...j-1}Q_{1..i-1}^{opt}(k)+Q(x_{i},x_{i+1},...,x_{j}),\label{bellmen-equation}
\end{equation}
where $Q_{1..i}^{opt}(k)$ denotes the optimal cumulative partial
score of partitioning the range $1...i$ into $k$ blocks.

\subsection{Scoring functions}
\label{Dynproscor} The scoring function $Q(B_{k})$ used in the dynamic
programming algorithm of data partition (\ref{cumulated-quality})-(\ref{bellmen-equation})
should be designed in such a way that it allows for
detection of the dense subgroups in the data. A scoring
function often used in the literature is the weighted sum of squares
of within block deviations of data points from mean

\begin{equation}
Q(B_{k})=\sum\limits _{n=i}^{n=j}\gamma_{k}\left[x_{n}-\frac{1}{j-i+1}\text{\ensuremath{\sum_{\nu=i}^{\nu=j}}}x_{\nu}\right]^{2}.\label{q-square}
\end{equation}
Often, weights $\gamma_{k}$ are assumed as normalizing factors for
numbers of elements in the block, which leads to the scoring function,
which we define as $Q_{1}(B_{k})$ 
\begin{equation}
Q_{1}(B_{k})=\frac{1}{(j-i+1)}\sum\limits _{n=i}^{n=j}\left[x_{n}-\frac{1}{j-i+1}\text{\ensuremath{\sum_{\nu=i}^{\nu=j}}}x_{\nu}\right]^{2}.\label{q1-square}
\end{equation}
The scoring function $Q_{1}(B_{k})$ is a within-block variance of
the data. Intuitively, $Q_{1}(B_{k})$ is a reasonable measure
of concentration of data points within the defined clusters for detecting
clusters in the data.

Other definitions of scoring functions are also possible and it seems
an interesting issue whether the use of other scoring functions can
improve data partitions, clusters detection and, consequently
estimation of the mixture parameters. Therefore, below we define other
scoring indexes, obtained by some modifications of $Q_{1}(B_{k})$
(\ref{q1-square}).

The second scoring function $Q_{2}(B_{k})$ is a within-block, standard
deviation 
\begin{equation}
Q_{2}(B_{k})=\sqrt{Q_{1}(B_{k})}.\label{q-std}
\end{equation}

We also use a third scoring function $Q_{3}(B_{k})$ defined as a
ratio of the within-block, standard deviation and the block sample
range, 
\begin{equation}
Q_{3}(B_{k})=\frac{\sqrt{Q_{1}(B_{k})}}{x_{j}-x_{i}}.\label{q-dimensionless}
\end{equation}
Intuitively, the scoring fucntion $Q_{3}(B_{k})$ takes smaller values
for blocks where data poins are concentrated close to the center and
larger values otherwise. The property of the scoring function $Q_{3}(B_{k})$
is that it is dimensionless and, in the large number of data points
limit, depends only on the shape of the probability distribution function
of the data.

Finally, we also introduce a fourth scoring function, $Q_{4}(B_{k},\Delta)$,
which is a modification of the scoring function $Q_{3}(B_{k})$, such
that some preference is given to wider blocks in comparison to narrower
blocks. The idea of giving preference to wider blocks is motivated
by the fact that very narrow blocks detected by $Q_{3}(B_{k})$
may correspond to random variations of the data rather than to the
true dense subgroups related to the structure of the Gaussian components.
In order to give some preference to wider blocks we modify $Q_{3}(B_{k})$
by introducing additional parameter $\Delta$, which leads to the
scoring function $Q_{4}(B_{k},\Delta)$ 
\begin{equation}
Q_{4}(B_{k},\Delta)=\frac{\Delta+\sqrt{Q_{1}(B_{k})}}{x_{j}-x_{i}}.\label{q-modified}
\end{equation}
Adding a positive constant $\Delta$ in the numerator of the above
expression results in limiting the possibility of ``shrinking''
the numerator of $Q_{4}(B_{k},\Delta)$ to values very close to $0$,
which can happen when narrow random ``peaks'' occur in the data.

\subsection{Scoring functions for binned data}
\label{Scorfun} For the case of application of the dynamic programming
method to binned data the appropriate modification of the expression
for the scoring function Q1 is 
\begin{equation}
Q_{1}(B_{k})=\sum\limits _{n=i}^{n=j}w_{n}\left(x_{n}-\sum_{\nu=i}^{\nu=j}x_{\nu}w_{\nu}\right)^{2},\label{Q1w}
\end{equation}
where $w_{n}$ is defined as in (\ref{defw}). Formulas for scoring
functions $Q_{2}$, $Q_{3}$ and $Q_{4}$, for binned data, are given
by (\ref{q-std}), (\ref{q-dimensionless}) and (\ref{q-modified}),
with (\ref{q1-square}) replaced by (\ref{Q1w}).

\subsection{Properties of different scoring functions}
\label{Propscor} Different scoring indexes may lead to different
partitions of the data. Therefore, in the computational experiments
further reported in this paper we apply and compare all scoring indexes
$Q_{1}(B_{k})$, $Q_{2}(B_{k})$, $Q_{3}(B_{k})$ and $Q_{4}(B_{k},\Delta)$.
Some preliminary observations (later systematically verified) are
summarized below.

The index $Q_{1}(B_{k})$ has a tendency to over-penalize wide components,
which can result in splitting some of the wider components into two
and in merging narrow components into one. The index $Q_{2}(B_{k})$
shows high sensitivity for the case where there is little overlap
between components. However, when the overlap between components increases,
it has the tendency to randomly merge components. The index $Q_{3}(B_{k})$
shows advantages following from its dimensionless construction and
often leads to correct partitions. However, it shows sensitivity
to noise in the data. Finally, the modified index $Q_{4}(B_{k},\Delta)$
allows for improving the performance of $Q_{3}(B_{k})$ by robustification
against noise in the data.

\section{Reference methods of setting initial condition for EM iterations }
\label{Refmet} We compare dynamic programming partitions to several
reference methods for setting initial conditions for EM iterations.
These methods were already studied in the literature [\cite{Biernacki2003,
Maitra2009, Fraley1999}] and they were
proven to be useful approaches for estimating mixture decompositions
of datasets. The first reference method of generation of initial
mixture parameters is the method of equal quantiles, used e.g. as
the default option in the software package Mclust [\cite{Fraley1999}].
Here bins are defined by equal quantiles of the dataset. Two
other reference methods are hierarchical clustering algorithms, e.g.
[\cite{Hastie2009}], where clusters of samples are created by successive
operation of merging based on distances between samples and/or distances
between clusters of samples. We apply two versions of hierarchical
clustering, with average and complete linkage [\cite{Hastie2009}]. Initial
values for component means, standard deviations and weights are defined
by blocks obtained by application of the hierarchical clustering
algorithms.

\section{Results}
\label{Compexper} We have conducted several computational experiments
for systematic comparisons of the methods of setting initial values
of parameters for EM iterations by the dynamic programming algorithm
and the three reference methods. We are using the following abbreviations
for algorithms for setting initial conditions: E-Q - equal quantiles
algorithm, H-clu-c - hierarchical clustering algorithm with complete
linkage, H-clu-a - hierarchical clustering algorithm with average
linkage, DP-Q1, DP-Q2, DP-Q3, DP-Q4($\Delta$) - dynamic programming
algorithm with scoring function Q1, Q2, Q3, Q4($\Delta$). In the
subsections below we first define performance criteria for comparing
results of applying different algorithms. Then we describe two groups
of computational experiments, artificially created data and proteomic
mass spectral data and we report results of comparisons of different
methods for setting initial conditions for EM iterations.

\subsection{Performance criteria}
\label{Perfcrit} Performance criteria for evaluating results of parameter
estimation algorithms are based on values of the differences between
true and estimated parameters or on the values of the log-likelihood
functions obtained in EM iterations, averaged over repeated experiments.
Since in our constructions of the quality criteria we aim at making
reasonable comparisons of different initialization algorithms for
datasets corresponding to different mixtures then we need to introduce
additional scalings and orderings of the values of differences or
log-likelihoods, as described below.

\paragraph{Difference between values of true and estimated parameters.}
The first approach to evaluating results of mixtures parameter estimation
algorithms is using a direct criterion given by a scaled difference
between estimated and true values. We use a scaled absolute difference
between true and estimated locations of components, averaged over
all components. Scaling is aimed at making the distribution of errors
invariant with respect to component widths and to components
weights. The~criterion is defined as~follows 
\begin{equation}
D=\frac{1}{K}\sum\limits _{i=1}^{K}\frac{\left\vert \mu_{i}^{\mathtt{true}}-\mu_{i}^{\mathtt{est}}\right\vert }{\sigma_{i}^{\mathtt{true}}}\sqrt{N\alpha_{i}^{\mathtt{true}}}.\label{q-criterion}
\end{equation}
In the above expression, $\mu_{i}^{\mathtt{true}}$, $\sigma_{i}^{\mathtt{true}}$
and $\alpha_{i}^{\mathtt{true}}$ are true parameters of the analyzed
mixture distribution, $K$ is the number of mixture components and
$N$ is the sample size. By~$\mu_{i}^{\mathtt{est}}$ we understand
the value of the estimated mixture component mean closest to $\mu_{i}^{\mathtt{true}}$.
Due~to skewness of the distributions of $D$, we use $\log(D)$ as
eventual measure of the quality of parameter estimation.

The value of the quality criterion $\log(D)$ averaged over
mixture datasets is denoted as 
\begin{equation}
Avg[\log(D)]\label{avg-q}
\end{equation}
and further used in reporting results of computational experiments.

\paragraph{Log-likelihoods.}
The direct criterion defined in the previous subsection can be used
only in the case where the true compositions of the analyzed mixtures
are known. Since we study both types of data with known and unknown
compositions of mixtures, we also use a second approach to evaluating
results of mixtures parameter estimation algorithms, based on values
of the log-likelihood functions. Values of the log-likelihoods obtained
in the EM iterations can be used for scoring performance of different
algorithms in both cases of known or unknown values of true mixture
parameters.

In the case of analysis of a single dataset on can use the obtained
values of log likelihoods to order performances of different algorithms.
Higher values of log likelihoods imply better quality of the solution
obtained by using an algorithm. There are exceptions from the rule
``higher likelihood $\rightarrow$ better estimate of mixture parameters''
caused by the possible existence of the spurious local maximizers
[\cite{McLachlan2000}], but their influence on obscuring results of
evaluations of performances of different algorithms is strongly reduced
by the used constraints on values of standard deviations of mixture
components.

Often we are not interested in using values of the log likelihood
functions for the analysis of one dataset but rather for comparisons
involving many datasets. In that case, typically the differences
between log-likelihoods obtained for different mixtures (different
datasets), are much larger than differences of log-likelihoods resulting
from applying different initialization algorithms for the same data-set.
Therefore orderings or scalings are used in order to compensate for
this. In this study we use the criterion applied previously in the
literature [\cite{Karlis2003, Biernacki2003}] defined
by the percentage (probability)\ of attaining ``maximum'' likelihood
by a method of initialization of EM iterations. By ``maximum'' likelihood
we understand the maximal value over all applied algorithms. We also
assume that ``a method no $m$ attained maximum likelihood'', means
that the difference between ``maximum'' likelihood and the $m$-th
likelihood is lower then $5\%$ of the range of all log likelihoods.
The value of this criterion, estimated by averaging over repeated
computational experiments, is denoted by 
\begin{equation}
Avg(P).\label{Probm}
\end{equation}

\subsection{Simulated datasets}
\label{Simdata} The first computational experiment involves analyses
of artificially created datasets, which are $10$ component Gaussian
mixtures, with known values of means, standard deviations and weights
of Gaussian components. In the simulated datasets we are controlling
the overlap between Gaussian components, due to its strong influence
on the results of the fit. Several measures of the degree of overlap between
two Gaussian components have been proposed in the literature, e.g., [\citet{Sun2011}].
They use different approaches. A simple method is based
on distances between Gaussian distributions (Mahalanobis,
Bhattacharyya). Here we define a simple parameter
$ov$ for measuring the degree of overlap between neighboring components,
\begin{equation}
ov=\exp\left[-\frac{\left\vert \mu_{i}-\mu_{i+1}\right\vert }{2\sqrt{\sigma_{i}^{2}+\sigma_{i+1}^{2}}}\right].\label{ov}
\end{equation}
Parameter $ov$ assumes value equal or close to zero for disjoint
components and larger values for components of stronger overlap. Maximal
possible value assumed by the overlap parameter is $ov=1$, which
occurs in the case where $\mu_{i}=\mu_{i+1}$. The definition of $ov$
in (\ref{ov}) can be interpreted as an adaptation/simplification
of the idea of the Bhattacharyya distance. The contruction (\ref{ov})
simplifies the overlap definition by the Bhattacharyya distance in
the sense that components with equal means, which show maximal possible
overlap $ov=1$, can be possibly distinguished based on the Bhattacharyya
distance by differences between variances. Despite this simplification,
the overlap measure (\ref{ov}) is useful for our analyzes, due to
the fact that we are not considering mixtures with components
of similar means and different variances (claw-like mixtures)[\cite{McLachlan2000}].

True parameters of each of the Gaussian mixtures are drawn randomly
in each stochastic simulation experiment. Draws of parameters of the
Gaussian mixtures are designed such that overlaps between neighboring
components are constant over one dataset. One stochastic simulation
experiment includes three steps: 
\begin{enumerate}
\item Draw randomly values of variances and weights of Gaussian components
$\sigma_{1},...,\sigma_{10}$ and $\alpha_{1},...,\alpha_{10}$. 
\item Define $\mu_{1}=0$ and compute values of means of Gaussian components,
$\mu_{2},...,\mu_{10}$ such that values of overlapping coefficient
(\ref{ov}) between successive components has a given constant value
$ov$. 
\item Generate 1000 independent samples of 10-component Gaussian mixtures
with parameters $\alpha_{1},...,\alpha_{10}$, $\mu_{1},...,\mu_{10}$
and $\sigma_{1},...,\sigma_{10}$. 
\end{enumerate}
Differences between stochastic simulation experiments involve different
methods for drawing weights $\alpha_{1},...,\alpha_{10}$ and standard
deviations $\sigma_{1},...,\sigma_{10}$ of Gaussian components in
the mixtures and different values of the overlapping coefficient $ov$.
Four groups of datasets are generated. Each includes 5~series of
experiments corresponding, respectively, to the following 5~values
of the overlapping coefficient: $ov=0.05,$ $0.1$, $0.15$, $0.2$,
$0.25$. Each series corresponds to one value of overlapping coefficient
$ov$ and includes $500$ datasets, generated according to steps
1-3. Different groups use different scenarios for generating weights
and variances of Gaussian components, described below.

\paragraph{Group 1: Equal mixing proportions. Low variability of standard deviations.}
In this group equal values of mixing proportions are assumed, $\alpha_{1}=\alpha_{2}=...=\alpha_{10}=0.1$
and values of component standard deviations are generated randomly
from uniform distribution $U(0.5,1)$ in each dataset. Values in
parenthesis give the range of possible changes of standard deviations.
So this scenario allows only for low (two-fold) variability of standard
deviations of Gaussian components.

\paragraph{Group 2: Equal mixing proportions. High variability of standard deviations.}
In this group again equal values of mixing proportions are assumed,
$\alpha_{1}=\alpha_{2}=...=\alpha_{10}=0.1$ and values of component
standard deviations are generated randomly from uniform distribution
$U(0.05,1)$. This scenario allows for high (20-fold) variability
of standard deviations of Gaussian components.

\paragraph{Group 3: Different mixing proportions. Low variability of standard
deviations.}
In this group different values of mixing proportions are assumed,
$\alpha_{1}=\frac{1}{55},\alpha_{2}=\frac{2}{55},...,\alpha_{10}=\frac{10}{55}$
and values of component standard deviations are generated randomly
from uniform distribution $U(0.5,1)$, which corresponds to low (2-fold)
variability of standard deviations of Gaussian components.

\paragraph{Group 4: Different mixing proportions. High variability of standard
deviations.}
In this group different values of mixing proportions are assumed,
$\alpha_{1}=\frac{1}{55},\alpha_{2}=\frac{2}{55},...,\alpha_{10}=\frac{10}{55}$
and values of component standard deviations are generated randomly
from uniform distribution $U(0.05,1)$, which corresponds to for high
(20-fold) variability of standard deviations of Gaussian components.

\paragraph{Comparisons of performances of different algorithms.}
In the computational experiments parameters of the Gaussian mixtures
were estimated by using EM iterations started with each of the above
described, seven algorithms of setting initial conditions, E-Q, H-clu-c,
H-clu-a, DP-Q1, DP-Q2, DP-Q3, DP-Q4($\Delta$). Performances of algorithms
are evaluated by using quality indexes $Avg[\log(D)]$ (\ref{avg-q}),
and $Avg(P)$ (\ref{Probm}). Results, obtained by averaging over
$500$ datasets in each of the data generation scenario are presented
in figure 1. This figure includes 5 subplots arranged in columns and
rows. Subplots in the two upper rows depict results of applying seven
algorithms of setting initial conditions in terms of $Avg[\log(D)]$.
Each of the subplots corresponds to one group of datasets (experiments).
In the top-left subplot, corresponding to the group 1 of experiments
with equal weights of components and low variability of standard deviations
of components, all initialization methods show high and similar performances.
In the subplot (second row, left column) corresponding to group 3
of experiments with low variability of standard deviations and different
component weights a method, which shows significantly lower performance
is the equal quantiles method E-Q. Other clustering methods (H-clu-c,
H-clu-a, DP-Q1, DP-Q2, DP-Q3, DP-Q4($\Delta$)) again show quite similar
performances, however differences are here bigger then in the previous
plot. In the subplots in the right column corresponding to groups
2 and 4 of stochastic simulation experiments, both with high variability
of standard deviations of Gaussian components, we can observe stronger
diversity between results of different initialization algorithms.
The performance of the equal quantiles algorithm E-Q is high in group
2 (equal component weights), however in group 4 (different component
weights) E-Q is the algorithm of the worst performance.

\begin{figure}
\includegraphics[width = 12.7cm]{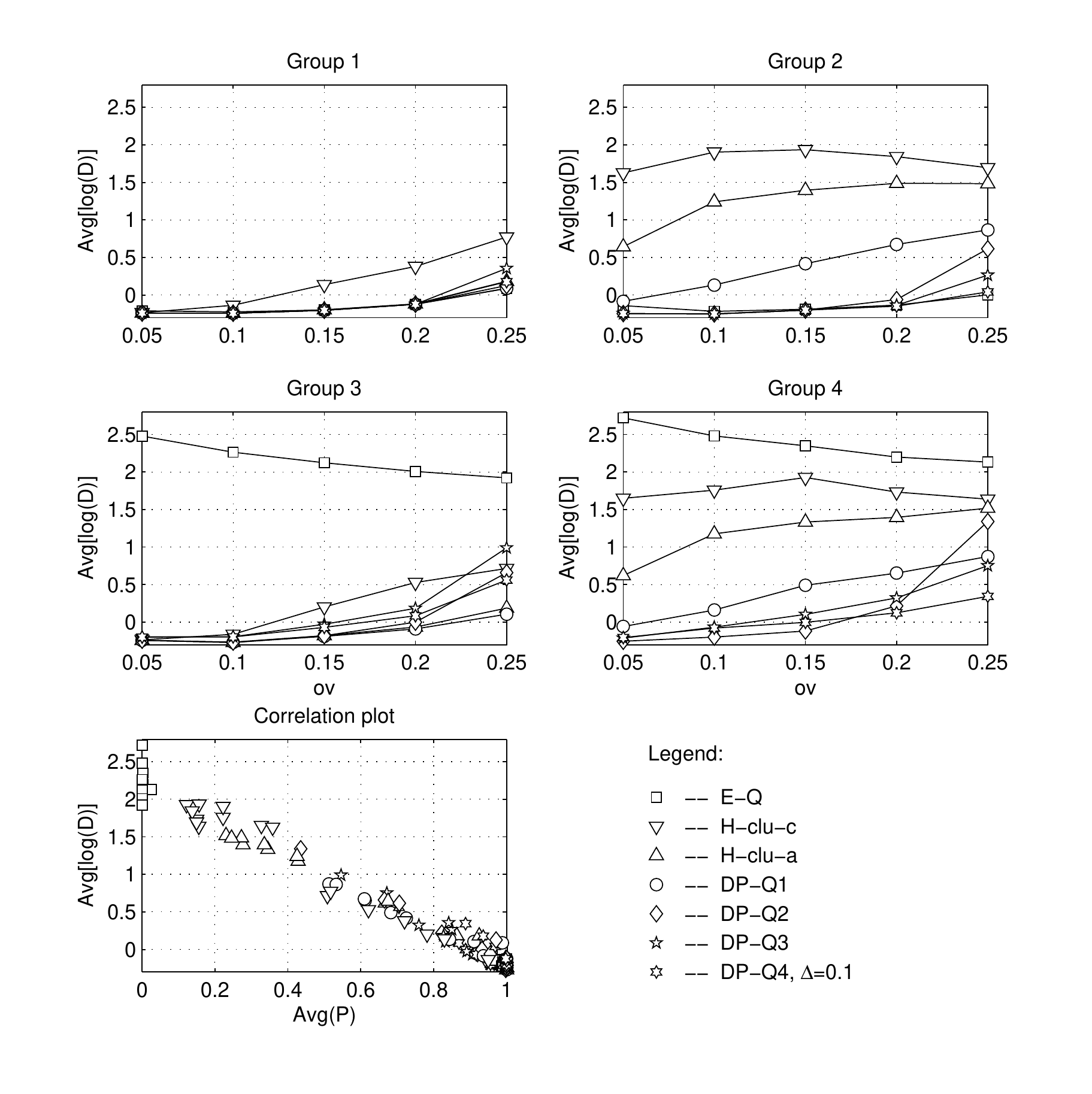}
\protect\caption{Comparisons of results of applying algorithms, E-Q, H-clu-c, H-clu-a,
DP-Q1, DP-Q2, DP-Q3, DP-Q4($\Delta$) for estimating mixture parameters
for simulated datasets. (Explanations in the text).}
\label{fig1}
\end{figure}

In all groups algorithms based on dynamic programming method DP-Q1,
DP-Q2, DP-Q3 and DP-Q4 either lead to similar values of $Avg[\log(D)]$
or outperform reference methods. The cases where performance of hierarchical
clustering method with average linkage can be slightly higher than
dynamic programming methods are groups 1 and 3 (low variability of
standard deviations). For groups 2 and 4 (high variability of standard
deviations) there is a strong advantage of the dynamic programming
algorithms over the reference methods.

High performance of the equal quantiles, E-Q, algorithm, in groups
1 and 2, can be considered as a kind of artifact. It does not follow
from the high capability of the algorithm to locate positions of components,
but rather from the fact that true values of component weights coincide
with equal quantiles assumed in the algorithm.

We can also make observations concerning comparisons between variants
of dynamic programming algorithms based on differetnt scoring functions.
Performance of the algorithm DP-Q1, based on the scoring function
(\ref{q1-square}) given by a sample variance, is high \ in
groups 1 and 3 where the variability of component standard deviations
is low. However, in groups 2 and 4 where the variability of component
standard deviations strongly increases, the algorithm DP-Q1 exhibits
worse performance compared to other variants of dynamic programming
method. This is consistent with the tendency of the dynamic programming
partition with the scoring function Q1 to incorrectly merge narrow
components. Performance of the algorithm DP-Q2 is high for low values
of the overlap coefficient $ov$, but strongly decreases with the
increase of $ov$. Performance of the algorithm DP-Q4($\Delta$),
$\Delta=0.1$ is better than the performance of DP-Q3.

In the bottom-left subplot we show a scatter-plot of values of indexes
$Avg[\log(D)]$ (\ref{avg-q})\ versus values of the probability
index $Avg(P)$ (\ref{Probm}). In the scatter-plot strong, negative
correlation between values of $Avg[\log(D)]$ and $Avg(P)$ is seen,
which confirms the potential of the index $Avg(P)$ to serve as a
reasonable estimate and a comparison tool for the performance of different
algorithms.

\subsection{Protein spectra dataset}
\label{Protspect} A proteomic mass spectrum contains information
about mass-to-charge (m/z) values of registered ions, denoted by $x_{n}$,
versus their abundances i.e., numbers of counts from the ion detector
denoted by $y_{n}$, $n$ denotes index of the data point. In real
experiments the dataset consists of more than one spectrum (sample).
To each point $x_{n}$ along the m/z axis correspond counts $y_{sn}$,
where $s$ denotes the index of the sample.

The second computational experiment in our study was the analysis
of the proteomic dataset, which included $52$ low resolution proteomic
mass spectra of human blood plasma [\cite{Pietrowska2011}]. This
dataset was obtained in the clinical study where blood samples were
collected in the group of 22 head and neck cancer patients and in
the group of 30 healthy donors. Raw spectra consisted of approximately
45~000 m/z values, covering the range of 2~000 to 12~000 Da. Spectral
signals were binned with the resolution 1 Da and baselines were removed
with the use of a version of the algorithm described in [\cite{Sauve2004}].
For further analysis only spectral fragments ranging from 2~000
to 4~120 Da have been selected. The choice of the range 2~000 to
4~120 Da was motivated by the fact that this fragment of m/z scale
contains several protein and peptide species interesting as potential
candidates for biomarkers.

\paragraph{Comparison of performances of different algorithms}
Computational experiments on the proteomic dataset involves modeling
spectral signals as mixtures of Gaussian components, estimating models
parameters by using EM algorithm and comparing qualities of models
obtained by using different algorithms of setting initial conditions
for EM iterations. The quality criterion for comparing different initialization
methods was $Avg(P)$, where averaging is done over all spectral signals
in the dataset. Clearly, the criterion $Avg[\log(D)]$ cannot be
used here due to the lack of knowledge on the true parameters of mixture
models.

For spectral signals with high variability of standard deviations,
strong, different overlaps between components and large number of
components, the differences between performances of different algorithms
of initialization of the EM iterations are magnified compared to the
simulated dataset. Application of algorithms, EQ, Hclu-c, Hclu-a,
DP-Q1, DP-Q2, DP-Q3, DP-Q4, for the proteomic dataset, lead to large
differences between values of log likelihood functions of models.
Since initialization methods EQ, Hclu-c, Hclu-a and DP-Q2 exhibit
significantly lower performance than DP-Q1, DP-Q3 and DP-Q4, then
we have confined the set of compared algorithms to those of the highest
quality, DP-Q1, DP-Q3 and DP-Q4. We have decomposed each of the spectral
signals into Gaussian mixture. EM algorithm was initialized with the
use of the following five algorithms: DP-Q1, DP-Q3, DP-Q4($\Delta=1$),
DP-Q4($\Delta=5$), DP-Q4($\Delta=10$). Decompositions were computed
with numbers of components $K$ ranging from 50 to 150.

\begin{figure}
\includegraphics[width = 12.7cm]{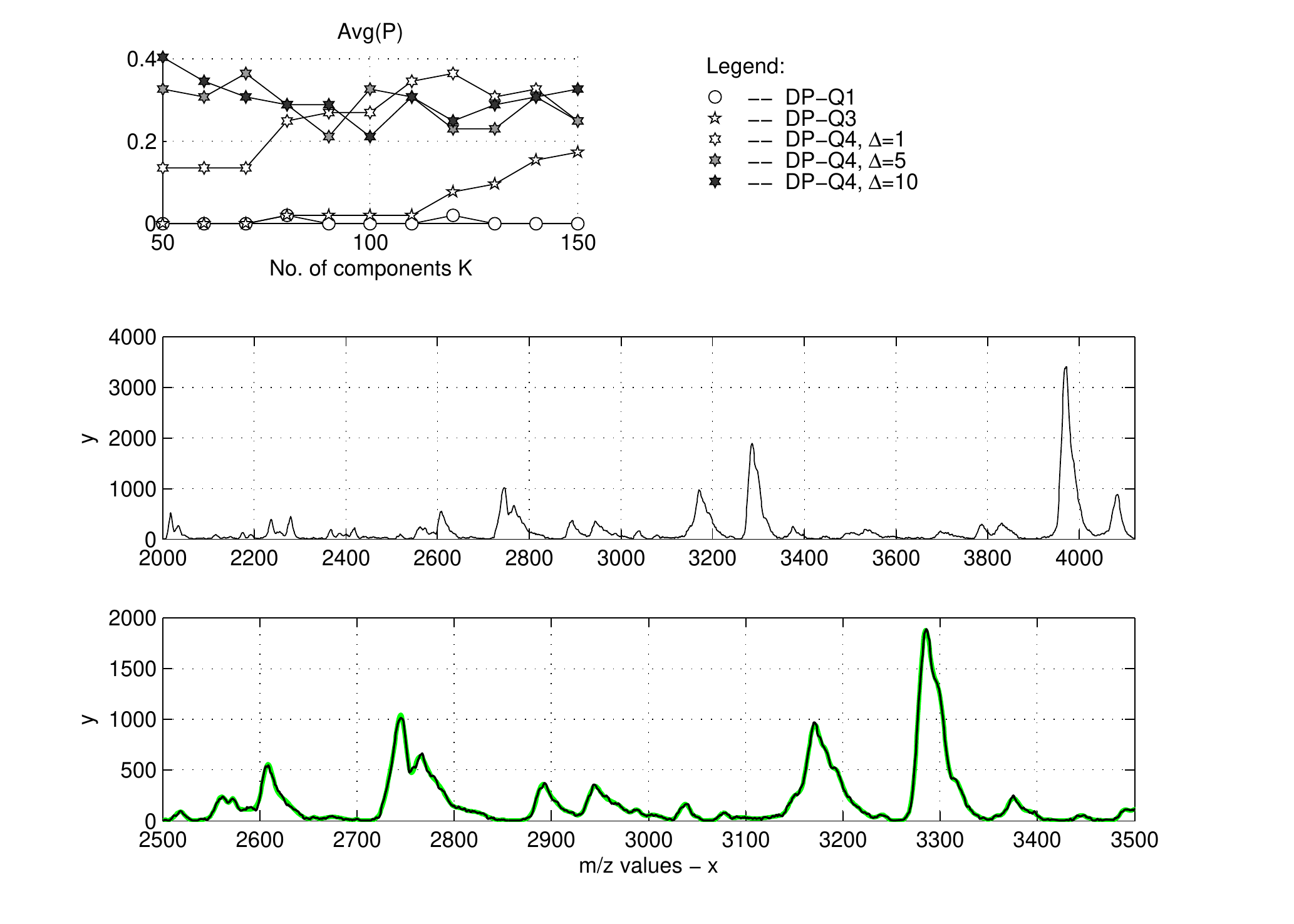}
\protect\caption{Upper panel: Comparison of performances of algorithms of setting initial
conditions DP-Q1, DP-Q3, DP-Q4($\Delta=1$), DP-Q4($\Delta=5$), DP-Q4($\Delta=10$)
for proteomic spectra dataset based on the $Avg(P)$ index. Middle
panel: Spectral signal $y_{n}=y_{1n}$. Lower panel: A fragment of
the spectrum $y_{n}=y_{1n}$ (black) versus its mixture model (green).}
\end{figure}

In figure 2, in the upper panel, we present the plot of values of
the index $Avg(P)$ versus the number of Gaussian components $K$.
One can observe that, on the average, DP-Q4 shows higher performance
than DP-Q1 and DP-Q3. High performance of the algorithm DP-Q4 was
observed for quite wide range of the parameter $\Delta$, $\Delta=1$,
$\Delta=5$, $\Delta=10$.

In the middle panel in figure 2 a plot of the spectral signal $y_{n}=y_{1n}$
(after preprocessing operations of baseline correction and binning)
corresponding to the sample no 1, within the range 2~000 to 4~120
Da. In the lower panel, the plot of a fragment of the spectrum $y_{n}=y_{1n}$
(black), within the range 2~500 to 3~500 Da, is drawn versus its
mixture model (green) obtained with the use of the algorithm DP-Q4
($\Delta=10$). The number of components ($K=90$) was estimated
with the use of Bayesian information criterion [\cite{Schwarz1978}]. One
can observe a good fit of the computed mixture model.

\section{Discussion}

Despite simplicity of construction of the EM algorithm, the research on its
performance and possible improvements is extensive and includes
several steps of the algorithm: initialization, stopping conditions, 
preventing divergence, execution [\cite{McLachlan2000}]. 
Modifications in each of the above steps interact one with another
in the sense of influencing the overall performance of the algorithm.  
In this paper we have focused on initialization of EM for certain types of mixture distributions.
We have also mentioned some solutions for stopping and 
preventing divergence. The latter of the above listed issues, concerning 
modifications of execution of EM steps for improving its performance, is
however also worth discussing due to its relations to the topic of our study.
 
Several papers introduce modifications of E and M steps of the EM algorithm
designed in such a way that searching through parameter space becomes more
intensive, which can help in avoiding local maxima of the likelihood
function and make the recursive process more robust against the choice of
initial conditions e.g., [\cite{Zhou2010}]. A group
of approaches to enhancing performance of EM iterations involves multiple
runs (threads) and repeats of EM steps combined with criteria of selecting
between threads [\cite{Karlis2003, Biernacki2003, OHagan2012}]. The~simplest
version of~short runs initiation [\cite{Karlis2003, Biernacki2003}] involves
generating multiple initial values for random methods and starting EM
iterations for the one corresponding to highest likelihood. Then only one
iteration process, namely the one, which attained the highest value of the
likelihood function is continued. A~recently developed implementation,
\textquotedblleft burn in EM\textquotedblright\ [\cite{OHagan2012}], involves
continuation of recursions of multiple threads combined with gradual
elimination of the worse on the basis of the value of likelihood function.
Several ideas of improving performance of EM iterations were
related to modifications of the log-likelihood function corresponding to the
Gaussian mixture model.One example of such an approach is the profile likelihood method in [\cite{Yao2010}]. 
Introducing constraints and/or modifications of the form of the
likelihood function both prevent divergence of iterations and lead to
improvement of performance of the corresponding variant of the EM algorithm.

Each of the above discussed approaches can be treated as competitive 
to our algorithm in the sense that it can lead to improvement of estimation
of parameters of mixture distributions. 
We did not present comparisons of our method to the above approaches.
However, according to out experience, for the type of data analyzed in this
paper, univariate, heteroscedastic, multi-component, precise initialization is 
more important than possible improvements following from modifications of 
execution of EM iterations. We should also mention that improvements of 
EM initialization can be combined with improvements in EM execution to lead 
to even better quality of mixture parameters estimation.

\section{Conclusions}
\label{Conclu} The first conclusion of our study is that initialization
methods based on dynamic programming, DP-Q1, DP-Q2, DP-Q3, DP-Q4($\Delta$),
show advantage over the reference methods EQ, Hclu-c, Hclu-a. We have
compared initialization algorithms for a variety of mixture data with
the overlap between neighboring components controlled by the parameter
(\ref{ov}) and different values of variances and component weights
(groups 1-4). This allowed for characterizing dependence of performances
of algorithm on parameters of mixture distributions. The advantage
of the dynamic programming initialization methods over the reference
methods is highest for heteroscedastic mixture distributions with
different mixing proportions.

The second conclusion is that performance of the dynamic programming
partitioning algorithm used for initialization of EM iterations depends
on the scoring function used in the algorithm. We have studied several
variants of the dynamic programming partition algorithms defined by
different scoring functions (\ref{q1-square})-(\ref{q-modified}).
The conclusion coming from these analyzes, drawn on the basis of both
$Avg[\log(D)]$ and $Avg(P)$ criterion was that for the type of datasets
with different mixing proportions and high variability of standard
deviations of components, the most efficient EM initialization method
is the dynamic programming algorithm with the scoring function Q4.

We have also applied the dynamic programming partition algorithms
DP-Q1, DP-Q3 and DP-Q4 for initialization of EM iterations for estimating
mixture model parameters for proteomic dataset including 52 low resolution
mass spectral signals. Comparisons of values of the $Avg(P)$ performance
criterion lead to the conclusion that again the method of the highest
efficiency is the dynamic programming partition with the scoring function
Q4.

The dynamic programming method with the scoring function Q4 needs
the adjustment of the parameter $\Delta$. However, computing
decompositions for several values of $\Delta$ ($1$, $5$, $10$)
leads to the conclusion that the algorithm shows high performance
for quite broad range of values of $\Delta$. So adjusting the value
of $\Delta$ can be done efficiently and robustly.

The dynamic programming algorithm applied to the partitioning problem
has a quadratic computational complexity with respect to the number
of elements of vector of observations $\mathbf{x}$. Despite computational
load, the advantage of using dynamic programming method for initialization
of the EM algorithm is the quality of the obtained mixture model.
In the majority of applications of mixture models the quality of the
model is more important than the computational complexity of the algorithm
used for the computations.

\section*{Acknowledgments}
This paper was partially financially supported by the scientific projects from the Polish National Center for Science (NCN) and the Polish National Center for Research and Development (NCBIR). JP was supported by NCN Harmonia grant DEC-2013/08/M/ST6/00924, AP was supported by NCN Opus grant UMO-2011/01/B/ST6/06868, MM was supported by NCBiR grant POIG.02.03.01-24-099/13. Computations were performed with the use of the infrastructure provided by the NCBIR POIG.02.03.01-24-099/13 grant: GeCONiI - Upper Silesian Center for Computational Science and Engineering. 

\section*{Supplementary materials}
Supplementary materials are Matlab scripts and functions for performing
comparisons of partitioning algorithms E-Q, H-clu-c, H-clu-a, DP-Q4
for the data described as Group 4 in section 6.2. Demo computations
are started by launching Matlab script partitions-em-demo. One stochastic
simulation experiment is performed (including three steps 1-3 listed
in section 6.2). Results of computations are shown by plots of partitions
and data histograms versus estimated probability density functions.
Values of errors $Avg[\log(D)]$ and likelihoods are also reported.
By modifications of the Matlab code other computational scenarios
for simulated data can be also easily realized.

\bibliography{Polanski_2015_ref}

\end{document}